\documentclass{ptephy}

\usepackage{amsmath}
\usepackage{amssymb}
\usepackage{graphicx}
\usepackage{dcolumn}
\usepackage{bm}

\begin{document}

\title{I${_2}$ molecule for neutrino mass spectroscopy: ab initio calculation of spectral rate}

\author{\name{Motomichi Tashiro}{1}\thanks{Author to whom correspondence should be addressed. E-mail: tashiro.motomichi@gmail.com}, \name{Masahiro Ehara}{1}, \name{Susumu Kuma}{2}, \name{Yuki Miyamoto}{3}, \name{Noboru Sasao}{2}, \name{Satoshi Uetake}{4}, and \name{Motohiko Yoshimura}{4}\thanks{Author to whom correspondence should be addressed. E-mail: yoshim@fphy.hep.okayama-u.ac.jp}}

\address{\affil{1}{Institute for Molecular Science, Nishigo-Naka 38, Myodaiji, Okazaki 444-8585, Japan}
\affil{2}{Research Core for Extreme Quantum World, Okayama University, Okayama 700-8530, Japan}
\affil{3}{Graduate School of Natural Science and Technology, Okayama University, Okayama 700-8530, Japan}
\affil{4}{Center of Quantum Universe, Okayama University, Okayama 700-8530, Japan}
}

\begin{abstract}%
It has recently been argued that
atoms and molecules may become good targets
of determining neutrino parameters still
undetermined, if atomic/molecular process is
enhanced by a new kind of coherence.
We compute photon energy
spectrum rate arising from coherent radiative neutrino
pair emission processes  of metastable
excited states of  I$_2$ and its iso-valent molecules,
$|Av \rangle \rightarrow |Xv' \rangle + \gamma + \nu_i\nu_j$ and 
$|A'v \rangle \rightarrow |Xv' \rangle + \gamma + \nu_i\nu_j$
with $\gamma$ an IR photon and $\nu_{i(j)}$ $i(j)-$th neutrino mass eigenstates,
and show how fundamental neutrino parameters may be determined.
Energies of electronically excited states of I$_2$, including the effect of 
spin-orbit couplings were calculated by the multiconfigurational second order perturbation (CASPT2) method. 
Summation over many vibrational levels of
intermediate states is fully incorporated.
Unlike atomic candidate of  a much larger energy difference such as Xe,
I$_2$ transitions from a vibrational level $A(v=0)$ to $X(v' = 24)$ give
an opportunity of determination of the mass type (Majorana vs Dirac distinction)
and determination of Majorana CPV (charge-conjugation parity
violating) phases, although the rate is much smaller.

\end{abstract}

\subjectindex{xxxx, xxx}

\maketitle

\section{Introduction}
Neutrinos are most common particles next to 3 K photons
in the present universe, yet their properties
eluded comprehensive experimental determination since
these neutral particles have only weak interaction.
The conventional target for exploration of neutrino properties
has been nuclei;
stable nuclei in neutrino oscillation experiments and
unstable nuclei  in other neutrino experiments.
Measured quantities derived so far by neutrino oscillation experiments 
are mass squared differences ($\Delta m_{ij}^2 \equiv m_i^2 - m_j^2$) 
and mixing angles summarized 
\cite{neutrino_2012_a}
by
\begin{eqnarray}
&&
 \Delta m_{21}^2 \sim 7.5 \times 10^{-5}\ \text{eV}^2 \,, \hspace{0.5cm}
|\Delta m_{31}^2| \sim 2.5 \times 10^{-3}\ \text{eV}^2
\,, 
\\ &&
\sin^2 \theta_{12} \sim 0.31\,, \hspace{0.5cm}
\sin^2 \theta_{23} \sim 0.42  \,, \hspace{0.5cm}
\sin^2 \theta_{13} \sim 0.024
\,.
\label{oscillation results}
\end{eqnarray}
It has been assumed in this analysis that there exist only three kinds of neutrino,
which we also follow throughout this work.
The other hint on the absolute mass scale is derived
from cosmological arguments, giving $\sum_i m_i < O(0.5)$ eV
\cite{cosmological_bound}.
The important quantities undetermined in oscillation experiments are
(1) absolute neutrino masses (impossible to determine in oscillation experiments),
(2) CP symmetry (C=charge conjugation, P = parity) violating phase
(CPV phase in short), and
(3) whether the massive neutrino belongs to the Majorana type \cite{majorana} or not.

There are two different kinds of ongoing experiments using unstable nuclei
for measurements of still undetermined neutrino parameters;
(1) the end point spectrum of beta decay of nuclei such as tritium for
the measurement of an averaged absolute neutrino mass value, 
and (2) the search of neutrino-less double beta decay 
for verification of lepton number violation related to
a finite Majorana  type of masses.
So far negative limits have been set in these experiments.
The reason for the use of  unstable nuclei is that the weak decay
rate such as the nuclear beta decay increases with a high positive power of 
the released energy, usually 5th power,
and the available nuclear energy of a few MeV 
gives a detectable rate.
But with the expected small neutrino mass of a fraction of eV
the energy mismatch becomes a serious problem, and
the determination of the absolute neutrino mass value
and exploration of undetermined important neutrino properties
are more and more difficult.

With the advent of remarkable technological innovations
manipulation  of 
atoms and molecules may contribute greatly to
fundamental physics.
Neutrino physics may also be one of these areas.
Atoms and molecules are target candidates of precision
neutrino mass spectroscopy, as recently emphasized in
Ref.~\cite{ptep_overview}, 
due to closeness of the energy released 
in their transition to expected neutrino masses.
The relevant process of our interest is cooperative
(and coherent, called macro-coherent subsequently) atomic de-excitation;
$|e \rangle \rightarrow |g\rangle + \gamma + \nu_i\nu_j$
where $\nu_{i(j)}, i(j) = 1,2,3$ is one of neutrino mass eigenstates.
$\gamma$ in the present work refers to a photon
in the visible to the infrared region.
The initial state $|e \rangle $ must be metastable, say its lifetime
$\gtrsim O(1)$ msec.
The process, as shown in Fig.~\ref{lambda-type atom for renp},
exists as a combined effect of second order in Quantum Electro-Dynamics (QED)
and weak interaction of the standard electroweak theory \cite{weak}. 

To obtain a measurable rate of the process,
it is crucial to develop the macro-coherence \cite{psr_dynamics, ptep_overview},
a new kind of coherence.
The macro-coherent emission of radiative neutrino pairs is stimulated by
two trigger irradiation of frequencies $\omega, \omega'$
constrained by $\omega + \omega' = \epsilon_{eg}/\hbar
\,, \omega < \omega'$, 
with $\epsilon_{eg} = \epsilon_e - \epsilon_g$ the energy difference
of initial and final states.
The measured photon energy in the de-excitation is given by the smaller frequency $\omega$.
The macro-coherence assures that the three-body process, $|e \rangle \rightarrow |g\rangle + \gamma + \nu_i\nu_j$,
conserves both the energy and the momentum.
Assuming that atoms in the states $|e\rangle$ and $|g\rangle$ can be taken infinitely heavy and
the atomic recoil may be ignored,
there exist threshold photon energies \cite{nu_observables_1} at
\begin{eqnarray}
&&
\hbar \omega_{ij} = \frac{\epsilon_{eg}}{2} - \frac{(m_i c^2 +m_j c^2)^2}{2\epsilon_{eg}}
\,.
\label{threshold location}
\end{eqnarray}
Each time the emitted photon energy decreases below
a fixed threshold energy of $\hbar \omega_{ij}$, 
a new continuous spectrum is opened, hence there are six energy thresholds
$\hbar \omega_{ij}\,, i,j = 1,2,3$ ($(i,j)$ threshold for brevity)
separated by finite photon energies.
Determination of the threshold location given by Eq.~(\ref{threshold location}), 
hence decomposition into six mass thresholds,
is made possible by precision of irradiated laser frequencies at 
$\omega \approx \omega_{ij}$ and
not by resolution of detected photon energy.

The macro-coherently amplified radiative emission of
neutrino pair has been called 
RENP (radiative emission of neutrino pair) \cite{ptep_overview},
which is the core idea of our neutrino mass spectroscopy
that may determine all unknown neutrino parameters.
It has been argued that this method
using atoms and molecules is ultimately
capable of determining the nature of neutrino
masses, Dirac vs Majorana distinction and
measuring the new Majorana source of CPV phases 
\cite{nu_observables_1,nu_observables_2}.
It is crucial for explanation of the matter-antimatter
asymmetry of the universe  to verify the Majorana nature of
neutrinos and determine CPV phases related to
the Majorana case \cite{fy-86, davidson-ibarra}.

There is an important constraint on possible target atoms/molecules
to obtain reasonable rates for realistic experiments.
The atomic operator involved in RENP has the character of M1$\times$E1
where the M1 operator (actually electron spin $\vec{S}$
in subsequent RENP formulas) governs the
weak interaction Hamiltonian of neutrino pair emission
and the E1 operator denotes the usual dipole interaction of QED.
The total angular momentum change via virtual intermediate state
requires that the $LS$ coupling scheme should be broken \cite{ptep_overview},
hence candidates should be sought in
heavy atoms/molecules.
This way we found Xe atom as a good candidate.

Xe atom is excellent, having a great discovery potential of RENP process, and
further expected to determine the absolute
neutrino mass scale and distinguish the normal mass vs the inverted mass
hierarchy (denoted by NH and IH respectively) \cite{ptep_overview}.
On the other hand, distinction of the mass type (Dirac vs Majorana)
requires a smaller released energy, of order a fraction of 1 eV,
as shown in Ref.~\cite{nu_observables_2}.
In the present work we shall study I$_2$ 
and its isovalent molecules for this purpose.

Molecules have a number of merits for RENP:
(1) homo-nuclear diatomic molecules such as I$_2$ have
vibrational states among which the usual E1 transition
(its Hamiltonian $\propto \vec{d}\cdot \vec{E}$)
is forbidden, (2) the richness of vibrational and rotational
levels makes it ideal to perform a systematic search of 
neutrino mass thresholds.
Moreover, I$_2$ molecule has a number of metastable
states with energies $\lesssim$ 1 eV above the electronically ground
vibrational excited levels.

Electronic excited states of I$_2$ molecule have been intensively investigated
by quantum chemical calculations \cite{ISI:A1971J783100039,ISI:A1994MW53500001,ISI:A1997YJ46400031}. 
The potential energy curves (PECs) and spectroscopic constants as well as transition properties 
of this system have been well examined, while the spin operator element which is relevant 
for the present neutrino mass spectroscopy has not been focused so much.
We shall present in this work RENP spectral rate using molecular
wave functions based on the first principle calculation.

We show in the present work
that the proposed I$_2$ de-excitation scheme gives
a chance of measurements, the Majorana vs Dirac distinction
and determination of CPV phases, which are,
in the quasi-degenerate case of neutrino masses, easier than NH vs IH
distinction at low photon energies where rates are largest.
A good sensitivity of the spectral shape to determination of the smallest neutrino mass of order
meV is also shown for this target molecule.

Throughout this work we assume that
the macro-coherent mechanism as proposed in Ref.~\cite{ptep_overview}.
In most of the rest except Sec.~\ref{SectionNameMolecularFactor}
where the atomic unit is used following the standard
practice of quantum chemistry, 
we use natural units
of $\hbar =1, c=1$.

\section{RENP amplitude and rate formula}

We shall recapitulate  from Ref.~\cite{ptep_overview} main features of the spectral formulas.

\begin{figure*}[htbp]
\centering\includegraphics[width=8.6cm]{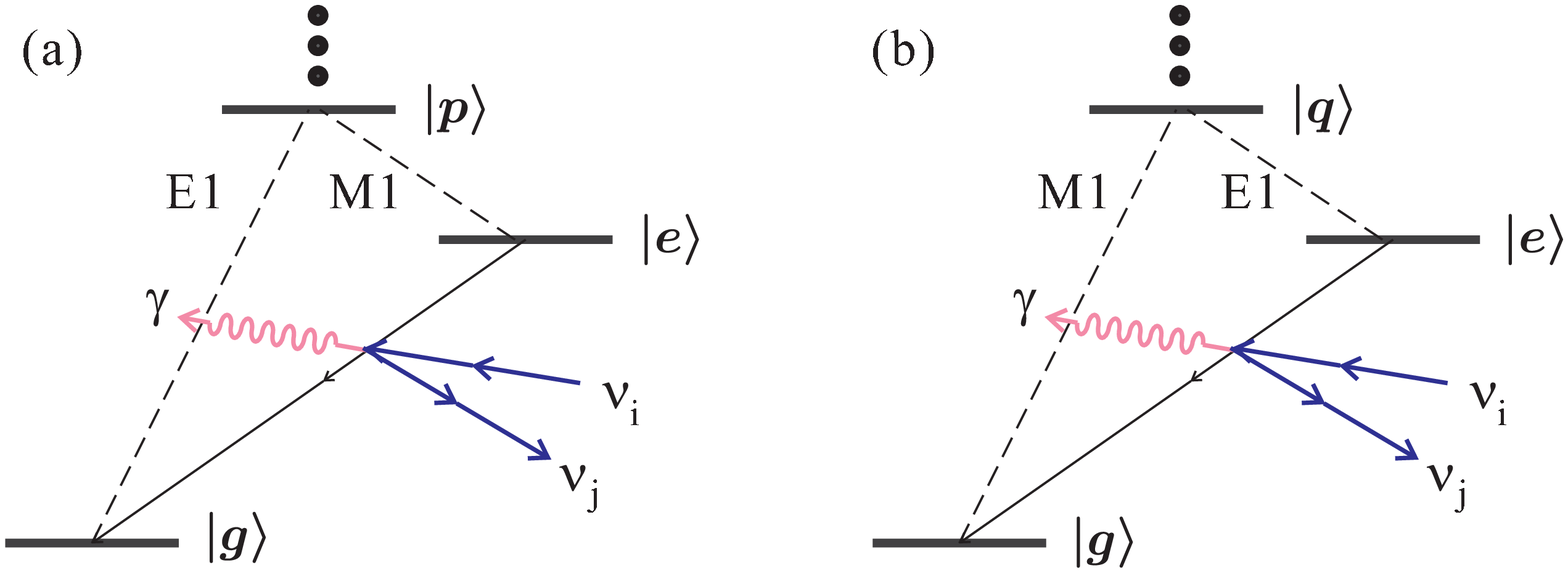}
   \caption{
Feynman-like diagrams  for  RENP from $\Lambda$-type atom/molecule,
$|e \rangle \rightarrow |g\rangle + \gamma + \nu_i\nu_j$,
with $\gamma$ a photon and $\nu_{i(j)}$ a neutrino mass eigenstate.
Virtual intermediate states $| p\rangle, |q\rangle$ should be summed over.
   Two-photon transition 
$|e \rangle \rightarrow |g\rangle + \gamma + \gamma$ 
may also occur via weak M1$\times$ E1 or E1$\times$M1 couplings to $|p\rangle$
or $|q \rangle$.
}
\label{lambda-type atom for renp} 
\end{figure*}

The amplitude corresponding to the Feynman-like diagram of 
Fig.~\ref{lambda-type atom for renp}
is given by the electroweak theory and reads as
\begin{eqnarray}
&&
\mathcal{M} =
G_F\vec{E}\cdot
\left(
\sum_p
\frac{ \langle g| \vec{d}|p \rangle  \langle p | \vec{S}| e\rangle}{\epsilon_{pg} -\omega}
+
\sum_q
\frac{\langle q | \vec{d}|e \rangle \langle g|\vec{S}_ |q \rangle}{\epsilon_{qg} -\omega}
\right)
\cdot\sum_{ij}a_{ij}\nu_j^{\dagger}\vec{\sigma}\nu_i 
\,, 
\label{renp amplitude}
\\ &&
a_{ij} = U_{ei}^*U_{ej} - \frac{1}{2} \delta_{ij}
\,,
\end{eqnarray}
where $U_{ei}$ (their relation to necessary mixing angles 
and CPV phases given in Eq.~(\ref{needed unitary n-matrix})) 
is the matrix element of neutrino mixing (expressions in terms
of measurable quantities given later), and
$\nu_i(\vec{p}, h)$ the neutrino plane wave function of momentum $\vec{p}$ and
helicity $h$ of mass $m_i$, $\vec{E}$ the electric field of irradiated
trigger laser of frequency $\omega < \epsilon_{eg}/2$, 
$\vec{S}$ the electron spin operator, and $\vec{d}$ the electric dipole operator.
Quantum numbers of states should include all of
electronic, vibrational, and rotational ones.
The sum over all vibrational modes of intermediate states, 
$|p\rangle$ and $|q\rangle$, is
particularly important for molecules.
For simplicity we ignore H\"onl-London factor 
\cite{hoenl-london} of order unity assuming
that rotational degrees of freedom are frozen.

Two terms in bracket of the right hand side of Eq.~(\ref{renp amplitude})
correspond to two different vertices,
weak M1 type of neutrino pair emission
(its Hamiltonian $\sim G_F \vec{S}\cdot \nu_j^{\dagger}\vec{\sigma}\nu_i $)
and QED E1 transition vertex in the second order perturbation theory
as depicted in Fig.~\ref{lambda-type atom for renp}:
the first term of molecular state
change of $|e\rangle \rightarrow |p \rangle \rightarrow |g\rangle $ shall be
designated by E1$\times$M1 and called (a) in the figure, and
the second term of molecular state
change of $|e\rangle \rightarrow |q \rangle \rightarrow |g\rangle $ shall be
designated by M1$\times$E1 and called (b).
Both of states $|p\rangle \,, |q\rangle$ are summed over, since
they are virtual intermediate states.

For isotropic medium without directional alignment by
a magnetic field, this amplitude squared gives the basic
rate formula (omitting the second contribution 
$C^b(\omega)$ from $|q\rangle$ in Eq.~(\ref{renp amplitude})),
\begin{eqnarray}
&&
\Gamma_{\gamma 2\nu}(\omega) = \Gamma_{dm} C^a(\omega) I(\omega)
\,,
\\ &&
\Gamma_{dm} = \frac{G_F^2 |\vec{E}|^2 n^2 V }{18\pi \bar{\epsilon}^2}
\,, 
\label{overall factor}
\\ &&
C^a(\omega) = \bar{\epsilon}\,^4
\sum_p \frac{\langle g| \vec{d} |p\rangle \cdot \langle p| \vec{d} |g\rangle
\langle e| \vec{S} |p\rangle \cdot \langle p| \vec{S} |e\rangle}{(\epsilon_{pg} - \omega)^2}
\,,
\label{molecular factor}
\\ &&
I(\omega) = \frac{1}{\bar{\epsilon}\,^2}
\sum_{ij}(B_{ij} I_{ij}(\omega) + \delta_M B^M_{ij} I^M_{ij}(\omega) ) 
\theta(\omega_{ij}-\omega)
\,,
\label{rnpe spectrum rate}
\\ &&
B_{ij} = | a_{ij}|^2
\,, \hspace{0.5cm}
B^M_{ij} = \Re (a_{ij}^2) 
\,,
\\ &&
\hspace*{-1cm}
I_{ij}(\omega) = \Delta_{ij}(\omega) \left(
\frac{1}{3}\epsilon_{eg}(\epsilon_{eg}-2\omega) 
+ \frac{1}{6}\omega^2
- \frac{1}{18}\omega^2 \Delta_{ij}^2(\omega) 
- \frac{1}{6}(m_i^2+ m_j^2)
- \frac{1}{6}\frac{(\epsilon_{eg}-\omega)^2}
{\epsilon_{eg}^2(\epsilon_{eg}-2\omega)^2}(m_i^2 - m_j^2)^2
\right)
\,,  
\nonumber \\ && 
\label{rnpe spectrum rate 1}
\\ &&
I^M_{ij}(\omega) 
= - m_i m_j \Delta_{ij}(\omega) 
\,,
\label{rnpe spectrum rate 2}
\\ &&
\Delta_{ij}(\omega) 
=\frac{1}{\epsilon_{eg} (\epsilon_{eg} -2\omega) }\left\{
\left( \epsilon_{eg} (\epsilon_{eg} -2\omega) - (m_i + m_j)^2\right)
\left( \epsilon_{eg} (\epsilon_{eg} -2\omega) - (m_i - m_j)^2\right)
\right\}^{1/2}
\,.
\label{rnpe spectrum rate 3}
\end{eqnarray}
In the overall rate factor $\Gamma_{dm}$ for diatomic molecules
the directionality
of trigger correlated with the molecular axis is taken into account by an extra $1/3$ reduction factor.
$\bar{\epsilon}$ is a reference energy, to make both $C^a(\omega)$ 
and $I(\omega)$ dimensionless.
In the rest of this work we take
$\bar{\epsilon} = \epsilon_{eg}$.
The function $\theta(x) = 0$ for $x<0$, $ = 1$ for $x > 0$
is the step function, giving rise to six mass thresholds in Eq.~(\ref{rnpe spectrum rate}).

$\delta_M =1$ for the case of Majorana neutrinos
and $\delta_M =0$ for the Dirac neutrino.
This term $\propto \delta_M$ 
exhibits the quantum mechanical interference
intrinsic to identical fermions of Majorana particles \cite{nu_observables_1},
which arises  from the anti-symmetric wave
functions of two identical fermions.
For the Dirac neutrino case,  emitted particles (except the photon) are a
neutrino and an anti-neutrino, two distinguishable particles,
and hence the interference terms are absent.

For stored field strength $|\vec{E}|^2$
we shall take its maximal value $\epsilon_{eg}n$ with $n$ the number
density of excited state targets.
More generally, this rate should be multiplied by the dynamical
factor $\eta_{\omega}(t)$ whose calculation requires
solution of the master equation given in Ref.~\cite{ptep_overview}.
It is important to keep in mind that
the medium polarization  $R_i\,, i=1,2$ between $|g\rangle $ and $|p\rangle $ states 
under trigger irradiation is contained
in the dynamical factor given by
\begin{eqnarray}
&&
\eta_{\omega}(t) = \frac{1}{L} \int_0^L dx \frac{|\vec{E}(x)|^2 |\langle p|(R_1- i R_2)|g\rangle (x)|^2}{4 \epsilon_{eg}n^3} 
\,,
\end{eqnarray}
an integrated quantity over the entire target at $0 \leq x \leq L$ along the
trigger irradiation.
Thus, a large macroscopic polarization is required for a large RENP rate.
The overall maximal rate in the unit of 1/time is 
\begin{eqnarray}
&&
\Gamma_{dm} = \frac{G_F^2 n^3 V }{18\pi \epsilon_{eg}}
\sim 26\ \text{kHz} \left( \frac{n}{10^{21}\ \text{cm}^{-3}} \right)^3 \frac{V}{10^2\ \text{cm}^3}
\frac{0.810\ \text{eV}}{\epsilon_{eg}}
\,.
\end{eqnarray}
The spectral information is given by $I(\omega)$
which is calculated using neutrino parameters 
experimentally determined, Eq.~(\ref{oscillation results}).
Calculation of the molecular factor $C(\omega)$
is the main subject of the next section.

\section{Molecular factor\label{SectionNameMolecularFactor}}

In this section, we investigate the possibility of the RENP experiment using I$_2$ molecule, with 
the electronically ground state $X\,0_g^+$ as the $|g \rangle$ state and the (metastable) 
electronically excited states $A\,1_u$ and $A'\,2_u$ as the $|e \rangle$ state. 
An advantage of the $A'\,2_u$ state over the $A\,1_u$ state is its longer natural lifetime.
In the simplified model for the RENP process $|A'v \rangle \rightarrow |Xv' \rangle + \gamma + \nu_i\nu_j$ in Ref.~\cite{ptep_overview}, only the I$_2$ $A\,1_u$ state is considered 
as the intermediate $|p \rangle$ state. However, other electronic states may have similar or larger 
contribution to the molecular factor.
We first calculate the molecular factor including several I$_2$ excited electronic states in the intermediate 
state summation, without considering vibrational states. There, we examine (1) which intermediate 
electronic state contributes the most significantly to the molecular factor, (2) relative weight of E1$\times$M1 and M1$\times$E1 
amplitude to the RENP process. 
Second, the effect of nuclear motion (vibrational states) on the molecular factor is investigated 
considering a single intermediate electronic state, in which we will examine how the molecular factor depends on $(A(v),X(v'))$ or $(A'(v),X(v'))$, a pair of the vibrational 
states on the $|e \rangle$ and $|g \rangle$ states, which may be useful in designing experiments in future.  
Finally, the magnitudes of the molecular factor are compared for isovalent molecules, I$_2$, Br$_2$ and Cl$_2$,
to study a dependence of the RENP rate on the spin-orbit couplings. 

In order to simulate realistic experimental situation, the I$_2$ energies including the spin-orbit effects are 
accurately calculated within the framework of the Born-Oppenheimer approximation. 

Throughout this section we use atomic units.
The conversion factor from the natural unit $C(\omega)$ to the atomic unit $C_{au}(\omega)$
is given by
\begin{eqnarray}
&&
C(\omega) = C_0 C_{au}(\omega) 
\,, \hspace{0.5cm}
C_0 = \epsilon_{eg}^4\frac{e^2 a_0^2}{E_H^2} \sim 3.9 \times 10^{-12}
\,,
\end{eqnarray}
with 
\(\:
a_0 \sim 0.53 \times 10^{-8}\ \text{cm}
\,, \:
E_H \sim 27\ \text{eV} 
\,, \:
e^2 = 4\pi \alpha \sim 4\pi/137
\,,
\:\)
and using $\epsilon_{eg} = 0.810$ eV.
This gives a rate unit $\Gamma_{dm} C_0 \sim 1.0 \times 10^{-7}$ Hz
for $n=10^{21}$ cm$^{-3}$ and $V=10^2$ cm$^3$.

\subsection{Detail of the calculation\label{subsection_detail}}

The electronic excited states of I$_2$ molecule were calculated by the 
multiconfigurational second-order perturbation method (CASPT2) 
\cite{ISI:A1996VM76500021,ISI:000085902300005}, 
based on the reference wavefunctions obtained by the state-averaged complete active space self-consistent field
(CASSCF) method \cite{ISI:A1985AGJ3300004,ISI:A1985AJG5000041}
with the atomic natural orbital-relativistic correlation consistent (ANO-RCC) all-electron basis set \cite{ISI:000220724300003}. 
The CASSCF method, using a linear combination of configuration state functions to describe electronic wavefunction, 
is adequate for calculation of electronically excited states with relatively small computational cost. 
More accurate energies, including dynamic electronic correlation, can be obtained by the CASPT2 method using 
the second order perturbation to the zero-th order CASSCF wavefunctions. 
The scalar relativistic effects were introduced at the CASSCF level by keeping the spin-independent (scalar) terms of the two-component 
reduced Hamiltonian using the 4th order Douglas-Kroll-Hess method \cite{ISI:000222680900005,ISI:000225440000011,ISI:000179042300015}. 
At this point, the spin-orbit effects are not included in the CASSCF wavefunctions and the CASPT2 energies. 
The spin-orbit effects were considered by the state-interaction method, where 
the spin-orbit matrix elements were evaluated between the CASSCF states 
using the Breit-Pauli operator \cite{ISI:000165193700015}, 
while the unperturbed diagonal energies were replaced by the CASPT2 energies. 
Final energies are obtained by diagonalization of this spin-orbit Hamiltonian. 
Our procedure is very similar to that employed in the calculation of the I$_2$ potential energy curves 
by Malmquist et al. \cite{ISI:000175958700011}.
The D$_{2h}$ symmetry was employed in our calculations. 
In the CASSCF calculation, total of 28 orbitals, 
1--7$a_g$(\textit{i.e.}~$1a_g,2a_g,\cdots, 7a_g$), 1--3$b_{3u}$,1--3$b_{2u}$,1$b_{1g}$,1--7$b_{1u}$,1--3$b_{2g}$,1--3$b_{3g}$ and 1$a_u$, were kept 
doubly occupied. In addition, the 1$a_g$ (1$\sigma_g$) and 1$b_{1u}$ (1$\sigma_u$) orbitals were 
frozen during the calculation. 
The remaining 50 electrons were distributed over 26 active orbitals: 
8--13$a_g$,4--6$b_{3u}$,4--6$b_{2u}$,2$b_{1g}$,8--13$b_{1u}$,4--6$b_{2g}$,4--6$b_{3g}$ and 2$a_u$. 
The state-averaging was performed over 18 electronic states: 
the three lowest $A_g$, one $B_{3u}$, one $B_{2u}$, one $B_{1g}$, one $B_{2g}$, one $B_{3g}$ and 
one $A_u$ states with singlet spin multiplicity, and one $B_{3u}$, one $B_{2u}$, one $B_{1g}$, 
three $B_{1u}$, one $B_{2g}$, one $B_{3g}$ and one $A_u$ states with triplet spin multiplicity. 
Note that these electronic states correlate with two iodine atoms in the ${}^{2}P$ state 
at the dissociation limit. 
By diagonalization of the spin-orbit Hamiltonian, total of 36 spin-orbit eigenstates were obtained.  
All these calculations were performed using the MOLPRO suite of programs \cite{MOLPRO}. 
Although the electric transition dipole moments between the spin-orbit eigenstates were obtained 
as part of the spin-orbit calculation in the MOLPRO programs,  
the matrix elements of the electronic spin operator $S$ were not provided. 
Thus, we explicitly evaluated the spin matrix elements between the spin-orbit eigenstates 
using the output of the eigenvectors. 

\begin{figure}[h]
\centering\includegraphics[scale=1.0]{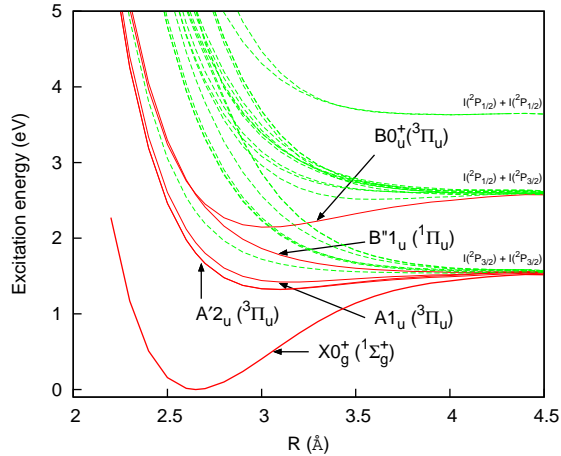}%
 \caption{
 Potential energy curves of I$_2$ electronic states. 
 The electronic states relevant to this work are indicated by solid red lines. 
 }
 \label{I2pecs}
\end{figure}

The potential energy curves of the calculated I$_2$ electronic states are shown in Fig.~\ref{I2pecs}. 
In the figure, the spin-orbit eigenstates are correlated with pairs of the atomic states 
I$({}^{2}P_{3/2})+$I$({}^{2}P_{3/2})$, I$({}^{2}P_{1/2})+$I$({}^{2}P_{3/2})$ or I$({}^{2}P_{1/2})+$I$({}^{2}P_{1/2})$, 
as the inter-nuclear distance becomes large.  
The dissociation and excitation energies of the electronic states relevant to this work,
as well as the vibrational energies on these states are summarized in Table \ref{tab1}.
For the purpose of this work, our results agree well with the experimental values. 

\begin{table}%
\caption{\label{tab1}
 Calculated vertical and adiabatic ($T_e$) excitation energies, dissociation energies ($D_e$) and vibrational energies ($\omega_e$) 
 for the I$_2$ $X$, $A'$, $A$, $B''$ and $B$ states. Experimental values are indicated in parentheses. 
$^a$Reference \cite{ISI:A1984SD76100008}.
$^b$Reference \cite{ISI:A1973P432700019}.
$^c$Reference \cite{ISI:A1982PR16400007}.
$^d$Reference \cite{ISI:A1981LN51200007}. 
$^e$Unbound in the calculated result. 
$^f$Reference \cite{ISI:A1985AHT6800017}. 
$^g$Reference \cite{ISI:A1982NP92500011}. 
$^h$Reference \cite{ISI:A1986A595600008}.
}
\begin{tabular}{ccccc}
\hline \hline
   State & Vertical excitation energies (eV) & $T_e$ (eV) & $D_e$ (eV) & $\omega_e$ (cm$^{-1}$) \\
\hline
${X}0_g^+ ({}^1\Sigma^+_g)$ &  0.0         &   0.0         &  1.53 (1.55)$^a$     &   220.1 (214.5)$^a$  \\
${A'}2_u ({}^3\Pi_u)$       & 1.88 (1.69)$^b$  &  1.32 (1.245)$^c$ &  0.21  (0.311)$^c$&   95.3 (108.8)$^c$  \\
${A}1_u ({}^3\Pi_u) $       & 1.94 (1.84)$^b$  &  1.42 (1.353)$^d$ &  0.11 (0.203)$^d$    &   77.5 (88.3)$^d$  \\
${B''}1_u ({}^1\Pi_u) $     & 2.57 (2.49)$^b$  &  $-$$^e$ (1.534)$^f$  &  $-$$^e$ (0.022)$^f$     &  $-$$^e$ (19.8)$^f$  \\
${B}0_u^+ ({}^3\Pi_u)$     & 2.59 (2.37)$^g$  &  2.14 (1.955)$^h$  &  0.44 (0.543)$^h$&  117.2 (125.7)$^h$  \\
\hline \hline
\end{tabular}
\end{table}

\subsection{Molecular factor with the fixed-nuclei approximation \label{subsection_molfacfixed}}

\begin{figure}[h]
\centering\includegraphics[scale=1.0]{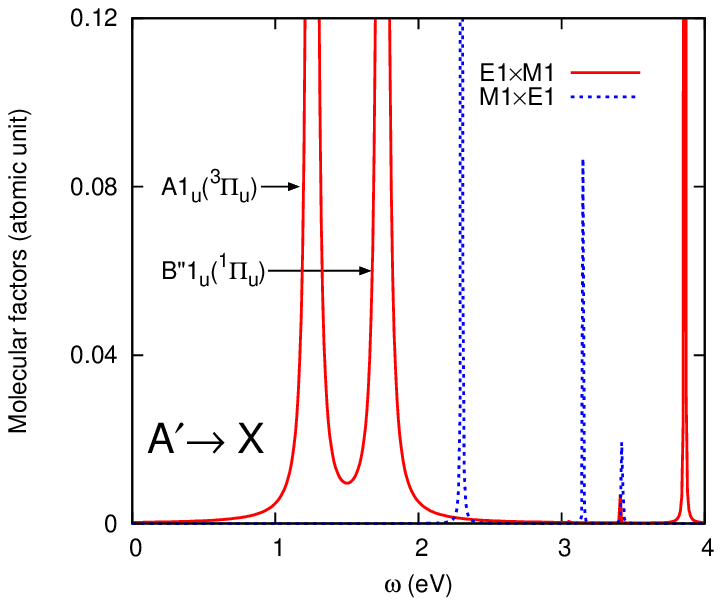}%
 \centering\includegraphics[scale=1.0]{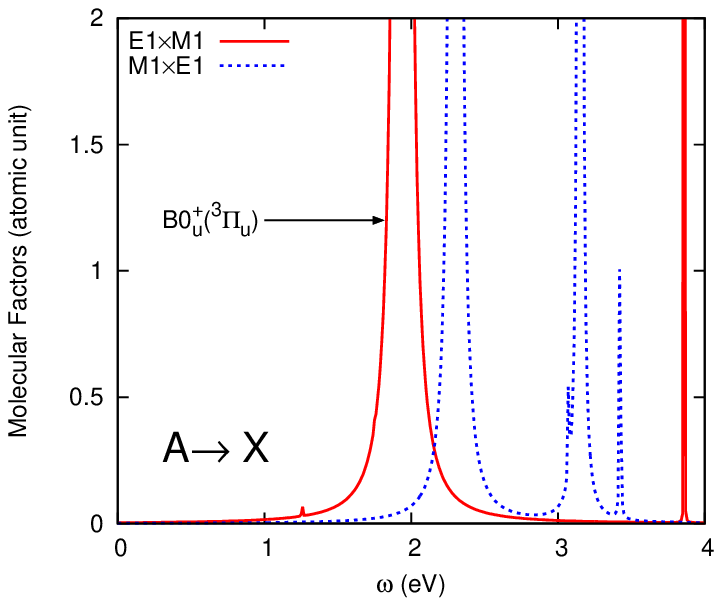}%
 \caption{
 Molecular factors evaluated at $R$ = 2.9 \AA. 
 The left panel: the $A'$ state was used as the $|e\rangle$ state. The right panel: the $A$ state was used as the $|e\rangle$ state. 
 The red line (E1$\times$M1, corresponding to Fig.~\ref{lambda-type atom for renp} (a)) represents molecular factor $C^{a}_{au} = C^{a}/C_0$  
 in Eq.~(\ref{molecular factor}). 
 The blue dashed line (M1$\times$E1, corresponding to Fig.~\ref{lambda-type atom for renp} (b)) 
 represents molecular factor $C^{b}_{au} = C^{b}/C_0$ in Eq.~(\ref{molfac_S}). 
 }
 \label{MF_fixed_R}
\end{figure}

In this work, we consider the I$_2$ $A\ 1_{u} ({}^{3}\Pi_{u})$ and $A'\ 2_{u} ({}^{3}\Pi_{u})$ states as $|e\rangle$ state and 
the $X 0_{g}^{+} ({}^{1}\Sigma^{+}_{g})$ ground state as $|g\rangle$ state, where electric dipole transition between 
$|e\rangle$ and $|g\rangle$ is forbidden or only weakly allowed. 
In order to inspect which I$_2$ electronic states contribute to the RENP rate as the intermediate 
$|p\rangle$ state, we calculated $C_{au}(\omega)$ at the fixed inter-nuclear distance of 2.9 \AA, which 
is located between the equilibrium points of the $X$ and $A$/$A'$ states, 
without considering rovibrational energies. 
All calculated spin-orbit eigenstates are included as intermediate 
$|p\rangle$ states in evaluating molecular factor $C_{au}(\omega)$. 
In Fig.~\ref{MF_fixed_R}, calculated molecular factor $C_{au}^a(\omega)$ corresponding to Fig.~\ref{lambda-type atom for renp} (a)
is shown 
along with the similar molecular factor $C_{au}^b(\omega)$ corresponding to Fig.~\ref{lambda-type atom for renp} (b), 
which has expression similar to Eq.~(\ref{molecular factor}) but with spin and dipole operators being swapped as

\begin{equation}
C^b(\omega) = \epsilon_{eg}^4
\sum_q \frac{\langle g| \vec{S} |q\rangle \cdot \langle q| \vec{S} |g\rangle
\langle e| \vec{d} |q\rangle \cdot \langle q| \vec{d} |e\rangle}{(\epsilon_{qg} - \omega)^2} .
\label{molfac_S}
\end{equation}

We only need energies around 0--0.5 eV in evaluating the RENP rate, although 
the molecular factors are plotted up to 5 eV in Fig.~\ref{MF_fixed_R} in order to inspect relative contribution of 
different intermediate electronic states. 
At low-energy region below 0.5 eV, the molecular factor for the E1$\times$M1 process is larger than 
that for the M1$\times$E1 process in both the $A' \to X$ and $A \to X$ cases.    
In other word, the first term in the bracket of Eq.~(\ref{renp amplitude}) dominates 
over the second term, which suggests that only the first term should be retained for the RENP using I$_2$ molecule. 
In the plots, we can observe several spike-like structures which represent contributions  
of the intermediate electronic states as labeled in the figure. 
When the $A\ 1_{u} ({}^{3}\Pi_{u})$ state is used as the $|e\rangle$ state, the $B\ 0_u^+ ({}^3\Pi_u)$ state predominates among contributions from the other states.
On the other hand, when the $A'\ 2_{u} ({}^{3}\Pi_{u})$ state is used as the $|e\rangle$ state, the $A\ 1_{u} ({}^{3}\Pi_{u})$ and $B''\ 1_{u} ({}^{1}\Pi_u)$ 
electronic states are especially important at low energy region relevant to this work, while the other states have 
negligible contributions.  

From here on, only the molecular factor in Eq.~(\ref{molecular factor}), which corresponds to the E1$\times$M1 process of 
Fig.~\ref{lambda-type atom for renp} (a), will be evaluated. 
For the $A \to X$ process, we will consider only the $B$ electronic state as the intermediate electronic state. 
For the $A' \to X$ process, the $A$ electronic state will be mainly considered as the intermediate electronic 
state, since the contribution of the $A$ state looks larger than the $B''$ state in the low energy region. 
Validity of this approximation, inclusion of only the $A$ electronic state in the $A' \to X$ process, will be inspected 
in the Appendix by comparing amplitudes evaluated by the $A$ intermediate electronic state and 
by the $B''$ intermediate electronic state. 

\subsection{Molecular factor with vibrational levels on the $X$, $A'$, $A$ and $B$ electronic states\label{subsection_molfacfXAA}}

\begin{figure}[h]
 \centering\includegraphics[scale=0.85]{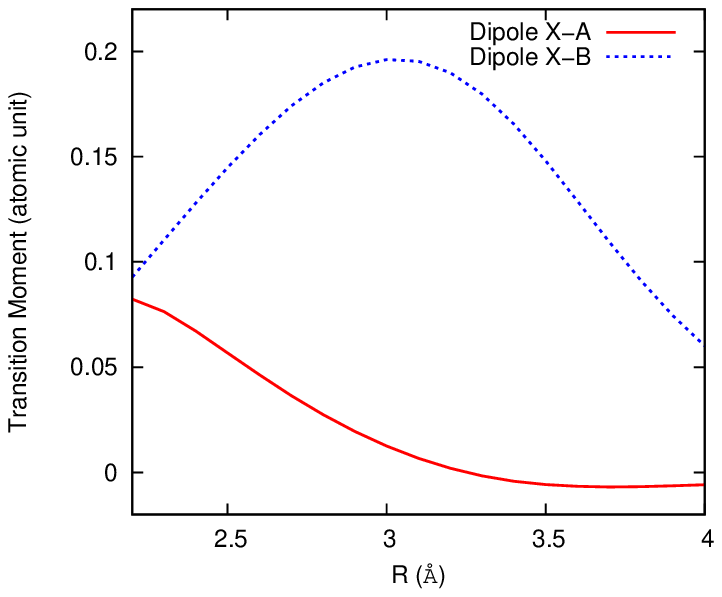}%
 \centering\includegraphics[scale=0.85]{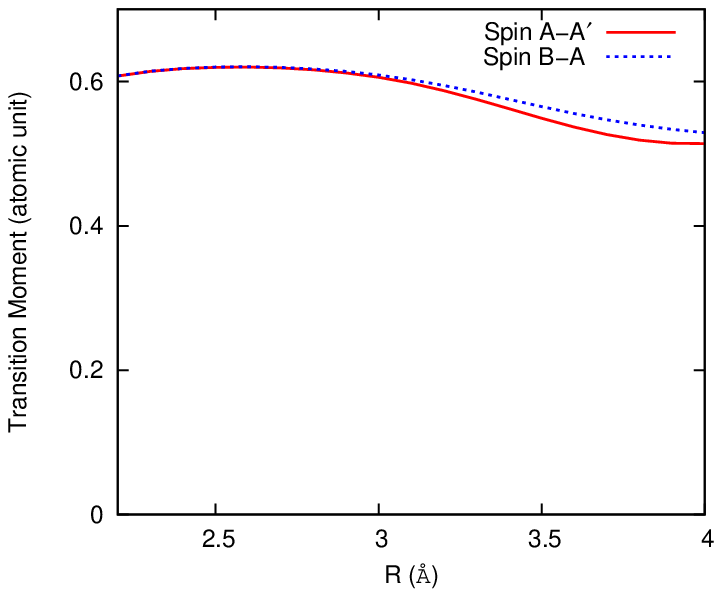}%
 \caption{
 Left panel: Electric dipole transition moment between the $X$ and $A$ states (red solid line), 
 and between the $X$ and $B$ states (blue dashed line). 
 Right panel: Spin transition moment between the $A'$ and $A$ states (red solid line), 
 and between the $A$ and $B$ states (blue dashed line). 
 }
 \label{XAAd_DTM_STM}
\end{figure}

\begin{figure}[h]
 \centering\includegraphics[scale=1.0]{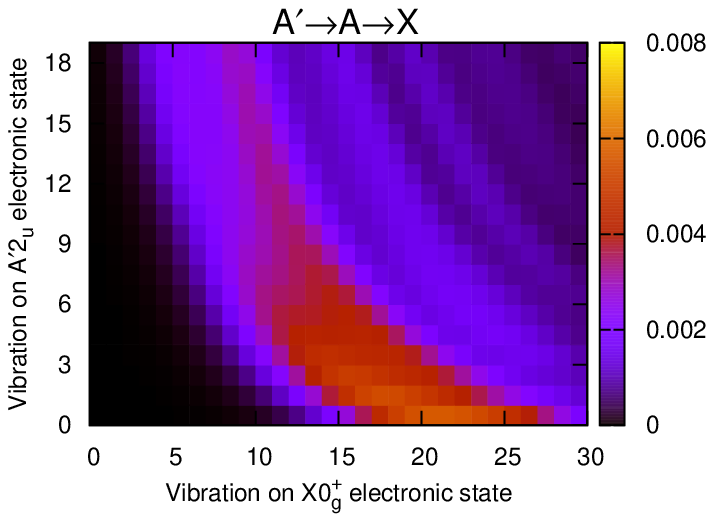}%
 \centering\includegraphics[scale=1.0]{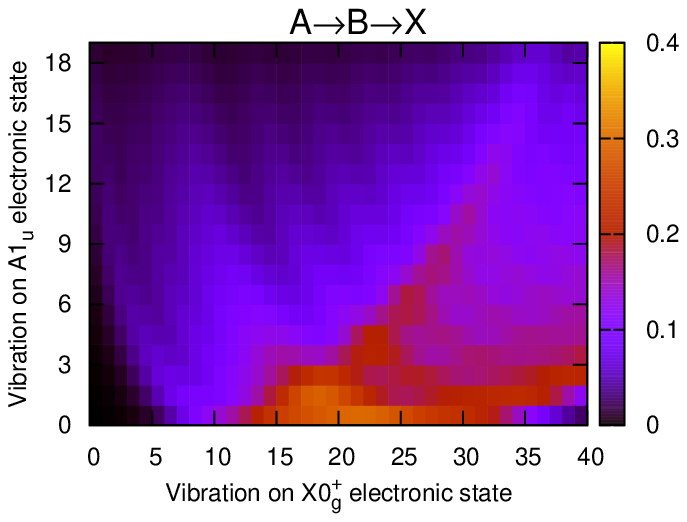}%
 \caption{
 Molecular factor $C_{au}^a(\omega)$ at $\omega$ = 0.4 eV, considering vibrational levels 
 on the $X$,$A'$,$A$ and $B$ electronic states.   
 The left panel represents the process in which the initial excited electronic state is the $A'$ state, 
 the intermediate electronic state is the $A$ state, and the final electronic state is the $X$ state. 
 The right panel represents the process where the initial excited electronic state is the $A$ state, 
 the intermediate electronic state is the $B$ state, and the final electronic state is the $X$ state.
 }
 \label{MF_vib_A_2D}
\end{figure}

\begin{figure}[h]
 \centering\includegraphics[scale=1.0]{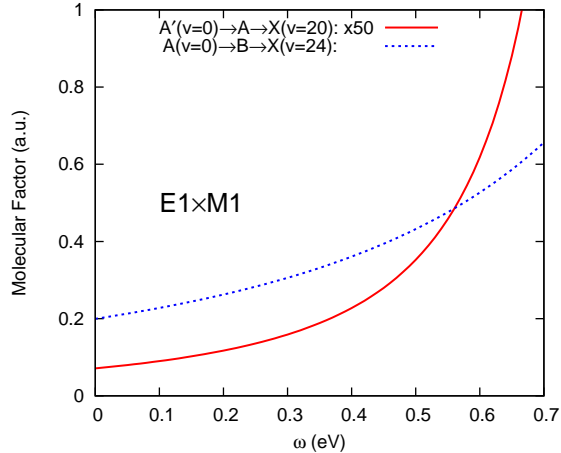}%
  \caption{
Molecular factor $C_{au}^a(\omega)$ as a function of photon energy $\omega$, considering vibrational levels 
 on the $X$,$A'$, $A$ and $B$ electronic states.
}
\label{MF_1D_A_B}
\end{figure}

Since the equilibrium inter-nuclear distances of the $X$ and $A'$ states, and of the $X$ and $A$ states 
are rather different, proper treatment of vibrational wavefunctions may be important to estimate the RENP rate. 
In this section, we evaluate the molecular factors considering vibrational levels 
on the $A'$, $A$ and $X$ electronic states for the $A' \to X$ process, and on the 
$A$, $B$ and $X$ electronic states for the $A \to X$ process. 
The ab initio energies of the $X$, $A'$, $A$ and $B$ electronic states, the same as shown in Fig.~\ref{I2pecs}, 
were fitted using the functional form taken from Ref.\ \cite{ISI:A1988R344300002}.  
The electric dipole transition moment between the $X$ and $A$ states and between the $X$ and $B$ states,  
the spin transition moment between the $A'$ and $A$ states and between the $A$ and $B$ states, 
shown in Fig.~\ref{XAAd_DTM_STM}, were fitted by polynomial functions. 
The vibrational energies and wavefunctions on each electronic state  
were obtained by direct diagonalization of the vibrational Hamiltonian using  
the pointwise coordinate representation of the wavefunctions, the discrete variable representation (DVR) 
method \cite{colbert1992}, with $R_{min}$ = 2.2 \AA, $R_{max}$ = 7.0 \AA~and $\Delta R$ = 0.006 \AA. 
Summation of the intermediate vibrational states, on the $A$ electronic state in the $A' \to X$ process, 
or on the $B$ electronic state in the $A \to X$ process, was taken up to $v$ = 40. 
These parameters for the DVR basis and the number of vibrational states in the summation were sufficient to 
obtain converged result. 

In Fig.~\ref{MF_vib_A_2D}, the calculated molecular factors at $\omega$ = 0.4 eV for the $A' \to A \to X$ 
process and for the the $A \to B \to X$ process are shown as functions of 
the vibrational levels on the initial ($A'$ or $A$) and the final ($X$) electronic states.  
In the $A' \to A \to X$ process shown in the left panel in Fig. \ref{MF_vib_A_2D}, the molecular factor is the largest around 
the point $(A'(v=0), X(v'=20))$. Starting from this point, the higher intensity region extends to the upper left direction. 
In the $A \to B \to X$ process shown in the right panel in Fig. \ref{MF_vib_A_2D}, 
the molecular factor is the largest around the point $(A(v=0), X(v'=24))$. 
In this case, the higher intensity region extends to the upper right direction. 
The structure of these higher intensity regions reflects the fact that 
the equilibrium points of the $X$ and $A'$ (or $A$) states are separated, and 
one of or both of the vibrational states on these electronic states should be 
sufficiently excited in order to achieve favorable overlap. 

In Fig. \ref{MF_1D_A_B}, the molecular factors $C_{au}^a(\omega)$ for the $A' \to A \to X$ and 
the $A \to B \to X$ processes are shown as a function of photon energy $w$. 
The vibrational level on the initial state, the $A'$ or $A$ electronic state, was 
selected to be $v = 0$, and the vibrational level on the final $X$ electronic state 
was selected to be $v = 20$ in the $A' \to A \to X$ case, and $v = 24$ in the $A \to B \to X$ case. 
As shown in the figure, the molecular factor for the $A \to B \to X$ process is 
about 50 times larger than that for the $A' \to A \to X$ process.

\subsection{Comparison with other molecules: Cl$_2$, Br$_2$ and O$_2$}

\begin{figure}[h]
 \centering\includegraphics[scale=1.0]{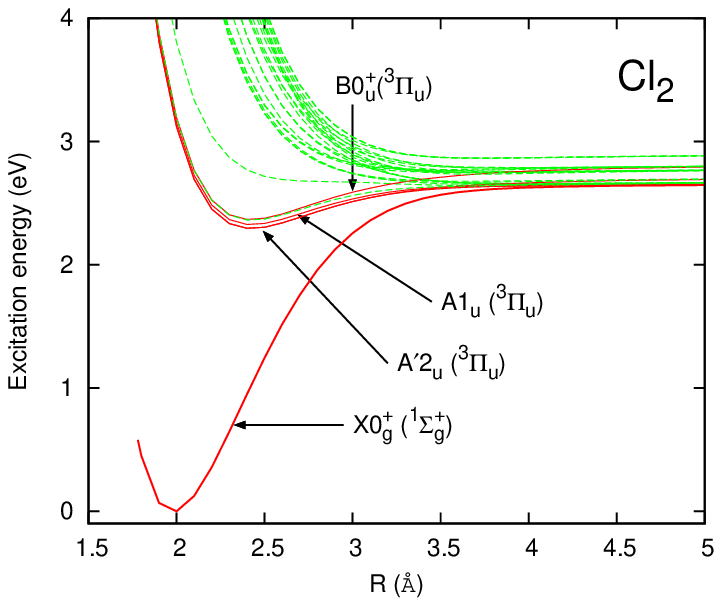}%
 \centering\includegraphics[scale=1.0]{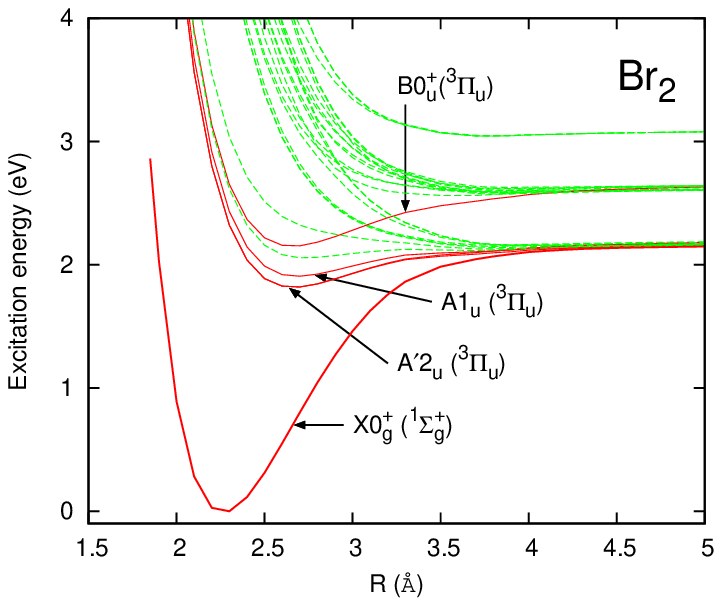}%
 \caption{
 Potential energy curves of Cl$_2$ (left panel) and Br$_2$ (right panel).
 }
 \label{Cl2Br2PEC}
\end{figure}

\begin{figure}[h]
 \centering\includegraphics[scale=1.0]{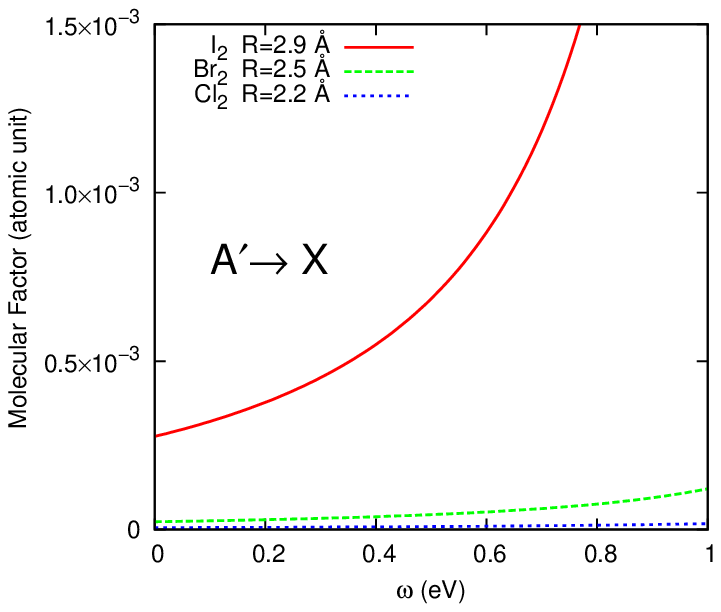}%
 \centering\includegraphics[scale=1.0]{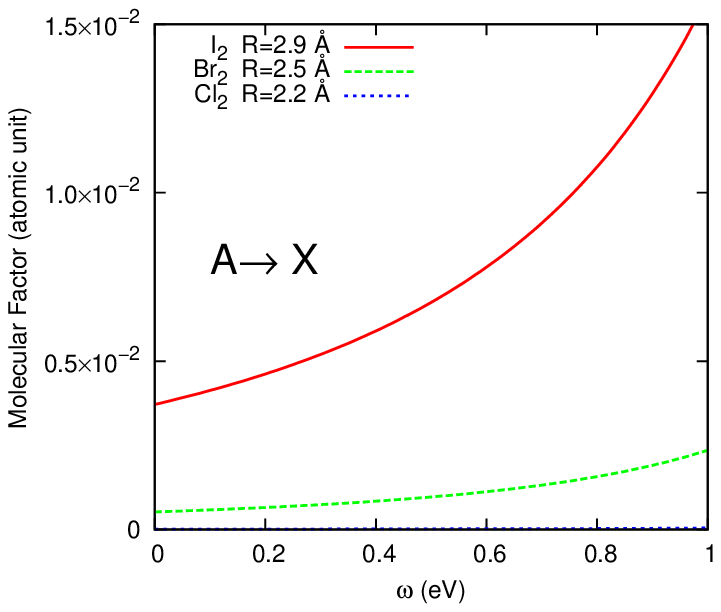}%
 \caption{
 Comparison of the molecular factors $C^{a}_{au}$ in Eq.~(\ref{molecular factor}) for I$_2$, Br$_2$ and Cl$_2$ molecules in the fixed-nuclei approximation. 
 The left panel: the $A'$ state was used as the $|e\rangle$ state. The right panel: the $A$ state was used as the $|e\rangle$ state. 
 The details of the calculation is the same as in Fig.~\ref{MF_fixed_R}. 
 }
 \label{I2_Br2_Cl2}
\end{figure}

In order to inspect the effect of atomic weight, or the strength of the spin-orbit couplings, 
we calculated the molecular factor $C^{a}_{au}$ for the E1$\times$M1 process for Cl$_2$ and Br$_2$ molecules with the fixed-nuclei approximation.  
The arrangements of the potential energy curves of Cl$_2$ and Br$_2$ are very similar to that of I$_2$, 
as shown in Fig.~\ref{Cl2Br2PEC}. In both Cl$_2$ and Br$_2$, the ground, first and second lowest excited states  
correspond to the ${X}0_g^+ ({}^1\Sigma^+_g)$, ${A'}2_u ({}^3\Pi_u)$ and ${A}1_u ({}^3\Pi_u) $ states, respectively, 
as in the I$_2$ case. Also, the location of the $B\ 0_u^+ ({}^3\Pi_u)$ state is similar to that of the $B$ state in I$_2$ molecule.  
The procedure of the calculation is the same as in the I$_2$ case described in Sec.~\ref{subsection_detail} and \ref{subsection_molfacfixed}.  
The molecular factors were evaluated for two different pathways as in the I$_2$ molecule. 
In the first case, we took the ${A'}2_u ({}^3\Pi_u)$ state as the initial state and the ${X}0_g^+ ({}^1\Sigma^+_g)$ as the final state. 
In the second case, the ${A}1_u ({}^3\Pi_u)$ state was taken as the initial state and the ${X}0_g^+ ({}^1\Sigma^+_g)$ was taken as the final state. 
All other 34 electronic states were considered in the summation of the intermediate state. 
We selected $R$ = 2.2 and 2.5 \AA ~for Cl$_2$ and Br$_2$, respectively, which are the middle points 
between the equilibrium points of the initial and the final states. 
In Fig.~\ref{I2_Br2_Cl2}, the molecular factors for I$_2$, Br$_2$ and Cl$_2$ are compared. 
The molecular factors for the $A \to X$ process are about 10 times larger than those for 
the $A' \to X$ process in all cases of I$_2$, Br$_2$ and Cl$_2$ molecules. 
The magnitude of the molecular factor is the largest for I$_2$, the smallest for Cl$_2$ in 
both the $A' \to X$ and $A \to X$ processes. 
These differences of magnitudes reflect the differences in strength of the spin-orbit couplings 
in these three molecules.

We also investigated the lighter molecule, O$_2$. 
In this case, the metastable $c{}^1\Sigma_u^-$ state was selected as the initial $|e\rangle$ state, 
and the $X{}^3\Sigma_g^-$ state as the final $|g\rangle$ state. 
Using the same procedure as we used for the I$_2$, Br$_2$ and Cl$_2$, the molecular factors for the E1$\times$M1 and 
M1$\times$E1 are calculated with the fixed-nuclei approximation. 
The molecular factor of O$_2$ for the E1$\times$M1 process is approximately $10^{-11}\sim10^{-10}$ in the energy range $\omega = 1 \sim 2$ eV, 
where the $A{}^3\Sigma_u^+$ state has the dominant contribution to the virtual state summation. 
For the M1$\times$E1 process, the molecular factor is about $10^{-7}$ in the energy range $\omega = 1 \sim 2$ eV, 
where the $1{}^1\Pi_g$ state has the dominant contribution to the virtual state summation. 
As expected from its small spin-orbit couplings, the molecular factor for O$_2$ molecule is smaller than the other I$_2$, Br$_2$ and Cl$_2$ 
molecules: the magnitudes of the molecular factors in the $A \to X$ process are $\sim10^{-2}$,$\sim10^{-3}$ and $\sim10^{-4}$, respectively.

\section{RENP spectral rate} 

A large macroscopic polarization necessary for significant RENP rates
is developed by two trigger laser irradiation of frequencies, $\omega$ and 
$\omega'$ with $ \omega +\omega' = \epsilon_{eg}$.
RENP amplitude is proportional to the polarization
$\langle g|(R_1 - iR_2)/2|p\rangle $ averaged over intermediate states $|p\rangle$.
Hence transition to a final single vibrational state $X(v')$ for $|g\rangle$
is selected out for the macro-coherently amplified RENP at each
experimental setup.

It is appropriate before detailed
presentation of numerical results to explain how experimentally
neutrino properties and parameters may be determined.
Suppose that two continuous wave
trigger lasers of frequencies, $\omega + \omega' = \epsilon_{eg}$,
are irradiated in counter-propagating
directions and two exciting pulse lasers of frequencies,
$\omega_P- \omega_S= \epsilon_{eg}$ (in order to
induce a Raman-type excitation $|g\rangle \rightarrow |e\rangle$)
are suddenly switched on. 
Only during this pulse irradiation RENP  occurs 
giving a unique signal; asymmetric directional increase of 
 light output of lower frequency $\omega$.
If statistics allows, one may hope to measure
parity-violating quantities such as emergence of
circular polarization from linear polarization.
Parity violating quantities that indicate unambiguous signal of
involved weak interaction appear with smaller rates,
at least by 100, than parity conserving quantities.
Observations at different
combinations of $(\omega, \omega')$ provide
spectrum rates at different $\omega$'s,
thus covering some range of frequencies that gives the
experimentally observed spectrum.
After this spectrum determination,
one compares the data with theoretical prediction computed
assuming some values of undetermined neutrino parameters and properties, and
finally the best fit with theoretical calculation determines these parameters.

The experimental method sketched here
is by no means unique, and one can think of other schemes.
Moreover, many simulations have to be done
to determine the signal level
once the best method of background rejection
is found.
It is suggested  that the
most serious background of the two photon
process may be rejected by
formation of condensed solitons \cite{ptep_overview},
which are a target state of coherent macroscopic target
entangled by static field condensates.

The quantity $C^a(\omega) I(\omega)/C_0 = C^a_{au}(\omega)I(\omega)$ 
in atomic units 
is plotted in the following figures except in the last three figures
where absolute rates are illustrated for
the target parameters of $n=10^{21}$ cm$^{-3}$ and $V=10^2$ cm$^3$.
We used for calculation of the
spectral rate numerical values of the mixing angles $\theta_{12}, \theta_{13}$,
as determined in neutrino
oscillation experiments Eq.~(\ref{oscillation results}),
thus giving numerical weights of six thresholds in Table \ref{tab_aij}.
(As usual, the conventional definition of oscillation angle factors
$c_{ij}= \cos \theta_{ij}\,, s_{ij}= \sin \theta_{ij}$ is used in this table.)
We also used mass constraints given by oscillation data Eq.~(\ref{oscillation results}).
The smallest mass defined by $m_0$ differs in the NH case 
where $m_0= m_1 (< m_2 < m_3)$
and in the IH case where $m_0= m_3 (<m_1 < m_2)$.
For large $m_0$, NH and IH differences are relatively small:
for instance, in the NH case $m_0= 100$~meV gives
three neutrino masses, $m_1=100, m_2=100.37$, and $m_3=111.8 $~meV's,
while in the IH case $m_0= 100$ meV gives $m_3=100, m_1=111.8$, and $m_2=112.1$~meV's,
their mass range within $\sim 10$ \%.
The mass pattern for a large $m_0$ has been called 
the quasi-degenerate case.

\vspace{0.5cm}
\begin{table}
\centering \caption{\label{tab_aij} 
The threshold weight 
$B_{ij}=|a_{ij}|^2=|U^*_{ei}U_{ej}-\delta_{ij}/2|^2$.}
 \begin{tabular}{cccccc}
  \hline \hline
   $B_{11}$& $B_{22}$ & $B_{33}$ & 
   $B_{12}+B_{21}$ & $B_{23}+B_{32}$ & $B_{31}+B_{13}$ \\
  \hline
   $(c_{12}^2c_{13}^2-1/2)^2$ & 
   $(s_{12}^2c_{13}^2-1/2)^2$ & 
   $(s_{13}^2-1/2)^2$ &
   $2c_{12}^2s_{12}^2c_{13}^4$ &
   $2s_{12}^2c_{13}^2s_{13}^2$ &
   $2c_{12}^2c_{13}^2s_{13}^2$ \\
  \hline       
   0.030 & 0.039 & 0.23 & 0.41 & 0.015 & 0.032 \\
  \hline \hline
   \end{tabular}
\end{table}

In all spectral rate figures
we take as the initial state the I$_2$ $A(v=0)$ state and as the final state
various vibrational states of $X(v')$.
It turns out that
numerically dominant contributions are to $X(v' = 22Ð-26)$.
Below we consider $X(v' = 24)$ as a representative example, since the molecular factors for $X(v' = 22Ð-26)$ are almost identical.
For simplicity we omitted contributions from intermediate
states other than the $B$ state, 
but included all numerically significant vibrational
states of $B(v'')$ as the intermediate state $|p\rangle$.
The RENP spectral rate from the metastable $A'$ state is calculated in a similar way.
The dominant contribution arises from $X(v' = 20)$.
However, the absolute rate is $\sim$ 50 times smaller than in the case of $A(v=0) \rightarrow X(v'=24)$.

\begin{figure*}[htbp]
\centering\includegraphics[width=8.0cm]{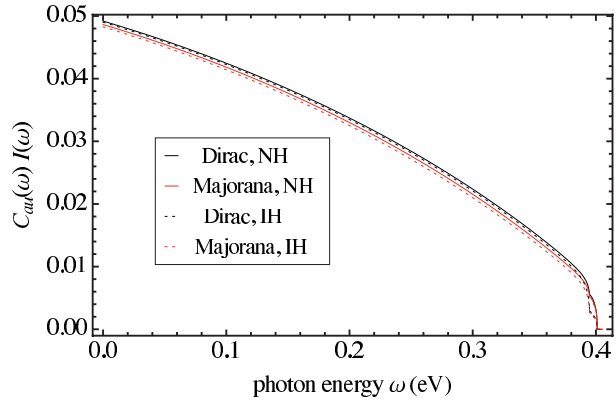}
   \caption{
I$_2$ spectrum 
$ C_{au}(\omega)I(\omega)$ 
taking $\bar{\epsilon} =\epsilon_{eg} = 0.810$ eV 
for the transition $A(v=0) \rightarrow X(v'=24)$.
   The Dirac NH case in solid black, Majorana NH case 
of $(\alpha, \beta') = (0,0)$ in solid red
 are compared with the IH cases in dashed colors,
   taking the smallest neutrino mass of 40 meV.
   }
   \label{i2 overall sepctrum}
\end{figure*}

In Fig.~\ref{i2 overall sepctrum} a global photon spectrum 
in the entire energy region is shown for
the transition $A(v=0) \rightarrow X(v'=24)$, where
four different cases are plotted, NH Dirac in black solid, NH Majorana
of $(\alpha, \beta-\delta) = (0,0)$ (the CP conserving case) in red solid,  
IH Dirac in black dashed, and
IH Majorana in red dashed (all with the smallest neutrino mass
of 40 meV) \cite{comparison_with_i2_ptep}.
All these cases appear nearly degenerate in the plot.

\begin{figure*}[htbp]
\centering\includegraphics[width=8.0cm]{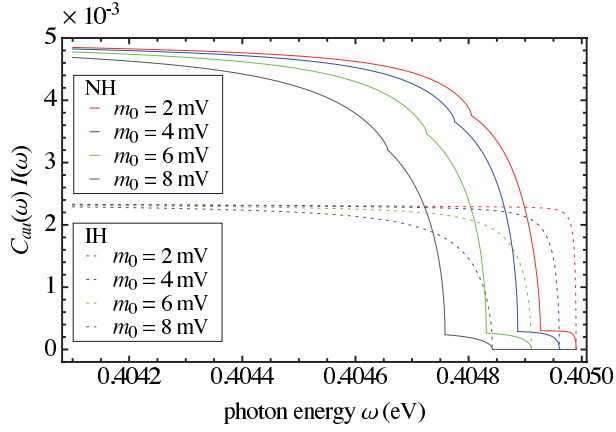}
   \caption{Sensitivity  of spectral shape to the smallest neutrino mass values,
   2,4,6,8 meV compared for the Majorana case of 
$(\alpha, \beta') = (0,0)$, NH  
(in solid) and IH (in dashed) cases. 
} 
\label{smallest neumass}
\end{figure*}

On the other hand, the enlarged spectrum in the threshold region is
shown for smaller mass values of $m_0$
in Fig.~\ref{smallest neumass}.
One can clearly observe  three kinks in the 
NH case which are identified as photon energy thresholds of
(11), (12), and (22).
The other three threshold kinks, (13), (23), and (33), 
in the NH case are further to the left
in this figure.
This figure suggests a good chance of determining the
smallest neutrino mass at the precision level of 1 meV,
if one has a large statistics data in the threshold
region.

We shall next examine the possibility of
Majorana-Dirac distinction along with determination
of CPV phases.
The relevant CPV phases $\alpha, \beta, \delta$
in the Majorana case appears in the matrix elements as
\begin{eqnarray}
&&
U_{e1}= \cos\theta_{13}\cos \theta_{12}
\,,
\hspace{0.5cm}
U_{e2}= \cos\theta_{13}\sin\theta_{12}e^{i\alpha} 
\,,
\hspace{0.5cm}
U_{e3}=\sin \theta_{13} e^{i(\beta -\delta)}
\,.
\label{needed unitary n-matrix}
\end{eqnarray}
In the Dirac case there is no CPV phase dependence of
the photon energy spectrum rate to this approximation.
The parameter $\delta$ alone
is accessible independently in neutrino oscillation experiments.
These phases appear multiplicatively, 
in the rate formula around three thresholds of (12), (13) and (23), 
as
\begin{eqnarray}
&&
\cos 2 \alpha
\,,
\hspace{0.5cm}
\cos 2 (\beta -\delta)
\,,
\hspace{0.5cm}
\cos 2 (\alpha - \beta + \delta)
\,,
\end{eqnarray}
which are further multiplied by the weight factor of Table \ref{tab_aij}
\cite{ptep_overview}
times a product of two masses $m_i m_j, i\neq j$.
There is thus no doubt that the Majorana-Dirac distinction
is easier for larger neutrino masses.
We shall introduce a new notation of CPV phase 
$\beta' \equiv \beta-\delta$ to simplify the formulas.
The case $(\alpha, \beta') = (0,0)$ corresponds to CP conserving 
(CPC) Majorana neutrino pair emission.

\begin{figure*}[htbp]
\centering\includegraphics[width=8.0cm]{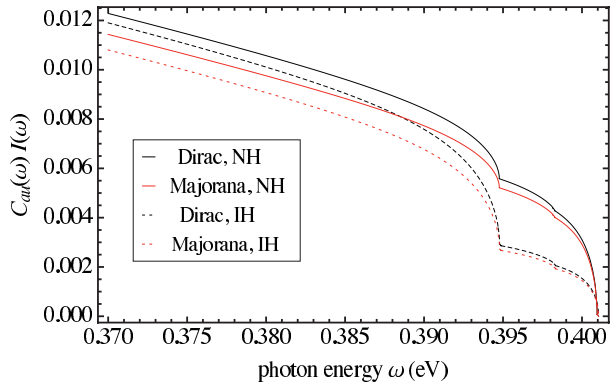}
   \caption{CP conserving Majorana case of 
$(\alpha, \beta') = (0,0)$
vs Dirac distinction, Dirac in black, Majorana in red.
   The smallest neutrino mass of 40 meV, and NH (in solid) and IH (in dashed) cases are assumed. 
} 
\label{md distinction}
\end{figure*}

As is evident in Fig.~\ref{md distinction}, 
the Majorana-Dirac distinction  in the proposed de-excitation of I$_2$ 
appears easier than NH-IH distinction at low enough photon energies
where rates are largest.
Numerically, the I$_2$ CPC Majorana rate for $m_0=40$ meV
is significantly different from the Dirac rate
by $\sim 0.07$ at a photon energy 0.37 eV,
while this difference for Xe $J=2$ transition
\cite{ptep_overview} never exceeds 0.2 \% at all photon energies.

\begin{figure*}[htbp]
\centering\includegraphics[width=8.0cm]{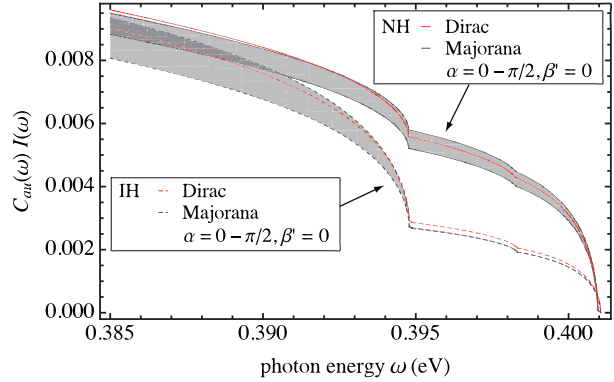}
   \caption{Majorana $\alpha$-band and Dirac photon spectrum
for $A(v=0) \rightarrow X(v'=24)$.
   Majorana band defined by variation of $\alpha$ in the range $0 \sim \pi/2$
   with a fixed $\beta'=0$.
   Dirac NH  is given in solid red and Dirac IH in dashed red, while
   the upper(lower) band corresponds to NH(IH).
    The smallest neutrino mass of 40 meV is assumed.
}
\label{maband vs d}
\end{figure*}

Study of the sensitivity to the CPV phase $\alpha, \beta'$
however requires a more careful analysis,
because the CPC case of $(\alpha, \beta')= (0,0)$
gives the largest destructive interference due to
the effect of identical Majorana fermions,
hence the Majorana-Dirac distinction is easiest in this case.
It is thus necessary to vary the CPV phases
$(\alpha, \beta')$ in their allowed range.
It turns out that with the given numerical
weight factors of Table \ref{tab_aij}
dependence of spectral rates on $\beta'$ is much weaker
than on $\alpha$. 
The most prominent kinks are at the (12) and (33) thresholds and
the highest sensitivity to $\alpha$ arises from the (12) threshold.
Under this circumstance we may introduce a useful concept
of Majorana $\alpha-$band which is defined by the region of
spectral rates, bounded by the largest Majorana rate at $\alpha = \pi/2$ 
and the smallest rate at $\alpha = 0$.
We illustrate this band structure in Fig.~\ref{maband vs d}.
Strictly, the $\alpha-$band is defined for
$\beta' = 0$, but dependence of the band shape
on $\beta'$ is small and we may approximately use the 
$\alpha-$band terminology for any value of $\beta'$.

\begin{figure*}[htbp]
\centering\includegraphics[width=8.0cm]{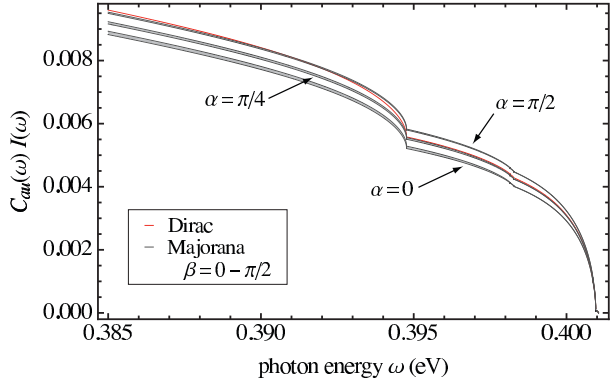}
   \caption{Majorana $\beta$-bands (shaded region bounded by black curves) for 3 values of
$\alpha = 0, \pi/4, \pi/2$ which are
well separated and Dirac (in red) photon spectrum
for $A(v=0) \rightarrow X(v'=24)$.
   A Majorana $\beta$-band is 
defined by variation of $\beta$ in the range $0 \sim \pi/2$
   with a fixed $\alpha$.
   NH case is shown.
      The smallest neutrino mass of 40 meV is assumed. 
} 
\label{mbband vs alpha}
\end{figure*}

On the other hand, variation of $\beta' $
gives much smaller band widths, and moreover
these bands are separated as $\alpha$ values vary,
as shown in Fig.~\ref{mbband vs alpha}.
It implies that experimental determination of $\beta' $
is much more difficult than $\alpha$
in the proposed I$_2$ de-excitation scheme.
We shall concentrate on determination of the CPV
parameter $\alpha$ in the following.

\begin{figure*}[htbp]
\centering\includegraphics[width=8.0cm]{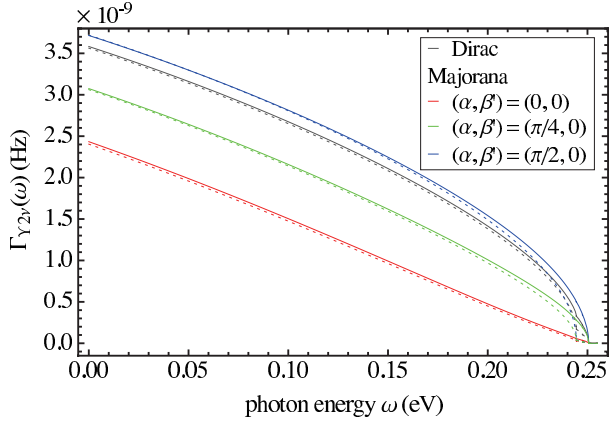}
   \caption{Three Majorana cases and Dirac case of the photon spectrum
for $A(v=0) \rightarrow X(v'=24)$.
NH in solid and IH in dashed.
Different colors correspond to Dirac and a few Majorana cases;
Dirac in black, Majorana of $(\alpha, \beta)=(0,0)$ in red,
Majorana of $(\alpha, \beta)=(\pi/2, 0 )$ in blue, 
and Majorana of $(\alpha, \beta)=(\pi/4,0)$ in green.
      The smallest neutrino mass of 250 meV is assumed.
      The transition $A(v=0) \rightarrow X(v'=25)$ gives  similar, but
      slightly smaller rate curves.
} 
\label{md nih xv20 250}
\end{figure*}

\begin{figure*}[htbp]
\centering\includegraphics[width=8.0cm]{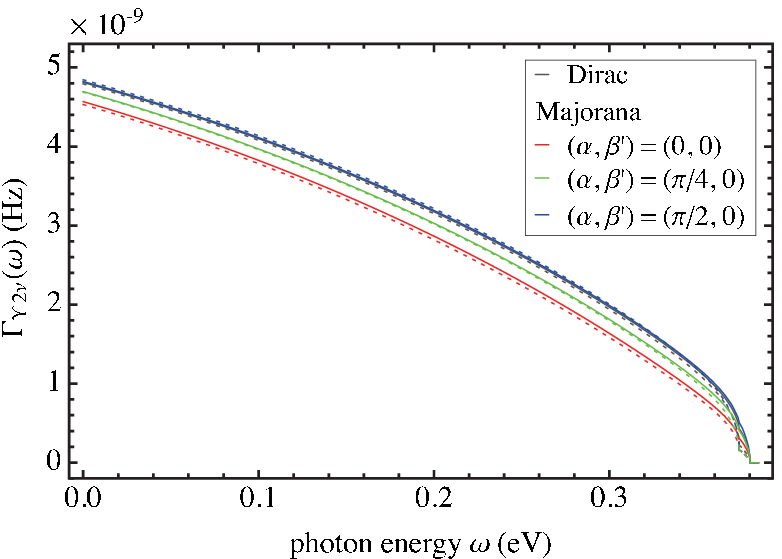}
   \caption{Similar plot to Fig.~\ref{md nih xv20 250}
   for $m_0=100$ meV and $A(v=0) \rightarrow X(v'=24)$. 
} 
\label{md nih xv20 100}
\end{figure*}

The spectral rate for a large value of the smallest neutrino mass, 250 meV
(a quasi-degenerate mass not excluded by
the cosmological bound \cite{cosmological_bound})   
is shown in Fig.~\ref{md nih xv20 250}, which
shows a great sensitivity to the Majorana-Dirac distinction
for this quasi-degenerate case.
Another example of spectral rate is shown for
$m_0= 100$ meV in Fig.~\ref{md nih xv20 100}
where the Dirac case and the Majorana case of $\alpha=\pi/2$
become nearly degenerate in the plot.
Thus, for relatively smaller mass values of $m_0$
one needs more detailed study, both by theoretical means
and numerical simulations,  for determination of the $\alpha$ parameter,
which is beyond the scope of the present work.

\begin{figure*}[htbp]
\centering\includegraphics[width=8.0cm]{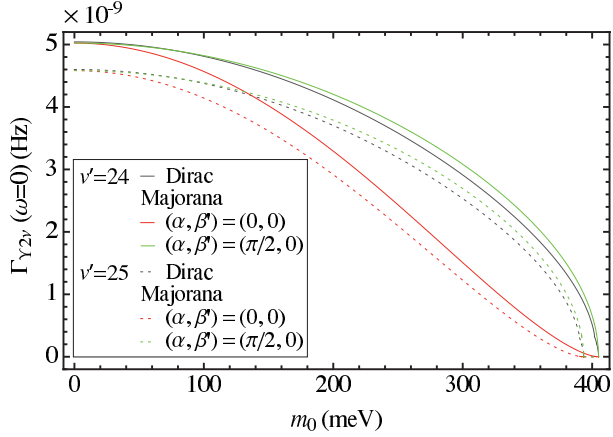}
   \caption{Zero energy limits of NH absolute spectrum rates
(in the unit of $10^{-11}$ Hz) are shown
for two transitions to $X(v'=24)$ (in solid)
and $X(v'=25)$ (in dashed). The Dirac cases are given in black, 
while the Majorana cases of $(\alpha,\beta)=(0,0)$ are given
in red and the $(\pi/2,0)$ cases are in 
green. 
The target parameters are  
$n=10^{21}\ \text{cm}^{-3}$ and $V= 10^2\ $cm$^3$.
$\epsilon_{eg} = 0.787$ eV was used for $X(v'=25)$. 
} 
\label{xv2120 md nh m0}
\end{figure*}

It is important to note that 
 comparison of two de-excitation to $X(v' = 24$ and 25)
offers a good tool of unambiguous experimental identification
of RENP (this can be done by a change of one of two
trigger lasers).
This is an advantage of molecular targets, since
one may compare two spectra of different transitions 
without much relying on the absolute rate scale.
The zero photon energy limit of absolute spectrum gives an overall
magnitude of the RENP rate, which then indicates relative difference of
these two different transitions. We plot in Fig.~\ref{xv2120 md nh m0}
these values for transitions to $X(v' = 24$ and 25) as functions
of the smallest neutrino mass $m_0$, which shows insensitivity
of rates for $m_0 < O(30)$ meV.
There is an interesting phenomenon of crossover of rates
occurring  at two points of the smallest mass $m_0 \sim 140$ meV 
and $\sim 390$ meV.
For the range of $\alpha$ parameter 
between these two values 
Dirac and Majorana $(\pi/2,0)$ rates  
to $X(v' = 25)$ of smaller atomic energy
separation overtake  Majorana $(0,0)$ 
rate to $X(v' = 24)$ of larger separation.
The simultaneous use of different vibrational transitions of molecules
may be effective for RENP identification
and exploration of the neutrino mass range as well.

The actual rate at each photon energy is equal to
the plotted values $C_{au}(\omega)I(\omega)$ multiplied by
the factor 
\begin{eqnarray}
&&
	\Gamma_{dm} C_0 \sim 1.0  \times 10^{-7}\ \text{Hz}
\left( \frac{n}{10^{21}\ \text{cm}^{-3}} \right)^3 \frac{V}{10^2\ \text{cm}^3}
\left( \frac{\epsilon_{eg}}{0.810\ \text{eV}} \right)^3
\,.
\end{eqnarray}
Note that we took $\eta_{\omega}(t) = 1$ which is however expected to be smaller than 1.

Let us compare I$_2$ RENP rate with Xe rate.
The I$_2$ rate is calculated as $\sim 5$ nHz at $\omega = 0$, assuming targets with $n=10^{21}$~cm$^{-3}$, $V=10^2$~cm$^3$, and $\eta_{\omega} = 1$.
Under the same experimental condition of targets,
the overall RENP rate for Xe atom is much larger, by 
$\sim10^5$.
The origin of this large rate difference
is explained as follows.
Dependence of the RENP rate (for definiteness at $\omega=0$) 
is roughly given by product of factors
as 
\begin{eqnarray}
\Gamma_{\gamma 2\nu}(0) \propto G_F^2 n^3V d^2 S^2
\frac{\epsilon_{eg}^3}{\epsilon_{pg}^2}
\,,
\end{eqnarray}
where we used $d$ and $S$ to mean the magnitudes of
dipole and the spin factor in the RENP 
amplitude.  Relation 
$\epsilon_{eg}\approx \epsilon_{pg}$ approximately holds both for Xe and I$_2$.
 An extra $n$-dependence 
in addition to the basic macro-coherence dependence
$\propto n^2 $ arises by
our assumption of taking the maximal RENP rate 
($\eta_{\omega}=1)$, with the stored field
energy density of $|\vec{E}|^2 = \epsilon_{eg}n$.
Difference of $\epsilon_{eg}$ between
Xe and I$_2$ is $\sim 10$ and the spin factor is
of the same order of unity for two targets.
The Franck-Condon like suppression factor 
\cite{fc_factor}
intrinsic to molecules is only 0.03 for I$_2$.
The rest of rate difference by $\sim 300$ is attributed to
difference in the transition dipole moment of Xe and I$_2$.
Unlike the simple dipole transition $6s \rightarrow 5p$ in Xe atom
the I$_2$ molecular dipole arises from 
$B(^3\Pi_u) \rightarrow X(^1\Sigma_g^+)$
involving different configurations and is much suppressed.
We definitely need solid environment of
the target number density of $O(10^{23})$ cm$^{-3}$ for realistic I$_2$ 
RENP experiments,
which is a challenge left to experimentalists.
A molecule closer to Xe with respect RENP appears to be
CsI.

\section{Summary}

An attractive feature of using molecules is that there are
many vibrational states available as a final molecular state for RENP which
makes easier experimental identification of the process.
We computed the RENP spectrum rate for homo-nuclear diatomic molecules
of isovalent series, Cl$_2$, Br$_2$ and I$_2$ and found that
the rate becomes larger as the atomic number increases, as is expected.
Even the largest I$_2$ rate is, however, much smaller than Xe atom rate.
The distinction of Dirac-Majorana neutrinos and the determination
of some range of the new CPV phase $\alpha$ are possible at photon energies
where rates are largest for the quasi-degenerate neutrino masses.
The  NH-IH distinction along with
the smallest mass measurement can be done around the threshold region.

\section*{Acknowledgment}
This research was partially supported by Grant-in-Aid for Scientific Research on Innovative Areas Extreme Quantum World Opened up by Atoms (Grant Nos. 21104002 and 21104003) from the Ministry of Education, Culture, Sports, Science, and Technology of Japan.
M.~T.~and M.~E.~thank to JSPS for financial support.

\appendix

\section{Appendix: Comparison of the contributions from the $A$ and $B''$ states in the I$_2$ $A' \to X$ process, 
including the effect of nuclear motion}

In the left panel of Fig.~\ref{MF_fixed_R}, with the fixed-nuclei approximation, the contribution of the $B''$ state to the molecular factor 
in the $A' \to X$ process looks close to that of the $A$ state at low energy region. 
Although we calculated the molecular factor for the $A' \to X$ process 
considering only the $A$ intermediate state in section \ref{subsection_molfacfXAA}, 
it is not clear if the $A$ state has really larger contribution for the molecular factor $C^{a}_{au}$ for the E1$\times$M1 process 
than the $B''$ state has, when the effect of nuclear motion is considered. 
Here, we evaluate the amplitude of the first term in the bracket of Eq.~(\ref{renp amplitude}), including nuclear motion, 
separately for the $A$ and $B''$ states, and estimate their contributions to the molecular factor. 
The reason for not treating molecular factor $C^a_{au}$ directly, but 
evaluating the first term in the bracket of Eq.~(\ref{renp amplitude}) is that the $B''$ state is repulsive, see Fig.~\ref{I2pecs}, 
and it is difficult to treat continuum nuclear wavefunctions in the expression of Eq.~(\ref{molecular factor}). 
The first term in the bracket of Eq.~(\ref{renp amplitude}), on the other hand, can be easily evaluated using the following 
expression:
\begin{equation}
\sum_{p} \frac{\langle g|d|p\rangle \langle p|S|e\rangle}{E_p - E_g -\omega} = 
\lim_{\epsilon \to +0} i \int_0^{\infty} dt e^{it\left(\omega+E_g\right)} \langle g|d e^{-it\left(\hat{H}-i\epsilon\right)}
S|e \rangle
\label{eq_wp}
\end{equation}
where $\hat{H}|p\rangle = E_p |p\rangle$ is assumed. 
This equation represents the Green's function by integral in time domain \cite{LevineBook}, 
and can be evaluated by the Fourier transform of the correlation function 
$\langle g|d e^{-it\left(\hat{H}-i\epsilon\right)}S|e \rangle$ obtained by solving time-dependent Schr\"odinger 
equation. 
For our purpose, the vibrational (nuclear) Hamiltonian on the $A$ or $B''$ electronic state is used for $\hat{H}$. 

\begin{figure}[h]
 \centering\includegraphics[scale=1.0]{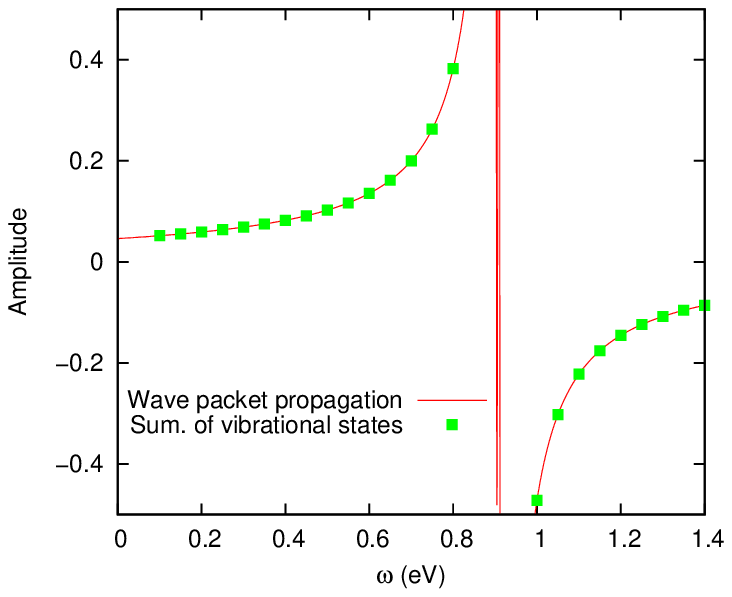}%
 \caption{ Comparison of Eq.~(\ref{eq_wp}) evaluated by the vibrational state summation ($v = 0 - 19$) and the 
 wave packet propagation on the I$_2$ $A1_u$ intermediate electronic state, using the I$_2$ 
 $A'(v=0)$ state as the initial state, and the $X (v'=20)$ state as the final state. 
 }
 \label{WP_VSUM_COMP}
\end{figure}

As a test, we evaluated the right hand side and the left hand side of Eq.~(\ref{eq_wp}) separately for the I$_2$ $A' \to A \to X$ 
process using explicit summation of the vibrational states, and the wavepacket propagation on the $A$ state, respectively. 
The same DVR basis parameters were used as in Sec.~\ref{subsection_molfacfXAA}.
The wave packet was propagated by expanding the time evolution operator in terms of 
Chebyshev polynomials \cite{ISI:A1984TS41400029}, with $\Delta t$ = 0.1 fs.  
Since many bound vibrational eigenstates contribute to the time evolution of the wavepacket, we need relatively long 
total propagation time, 7.0 ps. 
In addition, we had to leave small but finite value of $\epsilon$, $10^{-4}$ a.u., in order to 
converge the Fourier transform in the right hand side of Eq.~(\ref{eq_wp}). 
As shown in Fig. \ref{WP_VSUM_COMP}, numerical values of the left hand side and right hand side of 
Eq.~(\ref{eq_wp}) agree perfectly well in case of the $A$ electronic state. 

\begin{figure}[h]
 \centering\includegraphics[scale=0.85]{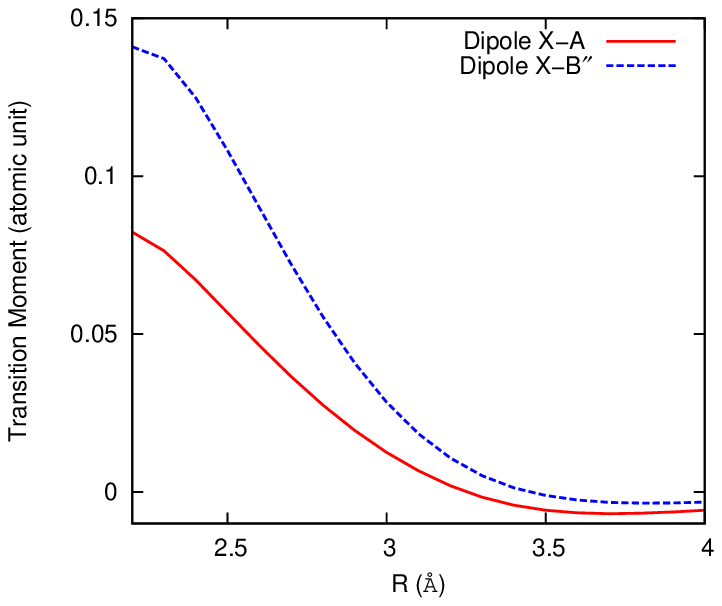}%
 \centering\includegraphics[scale=0.85]{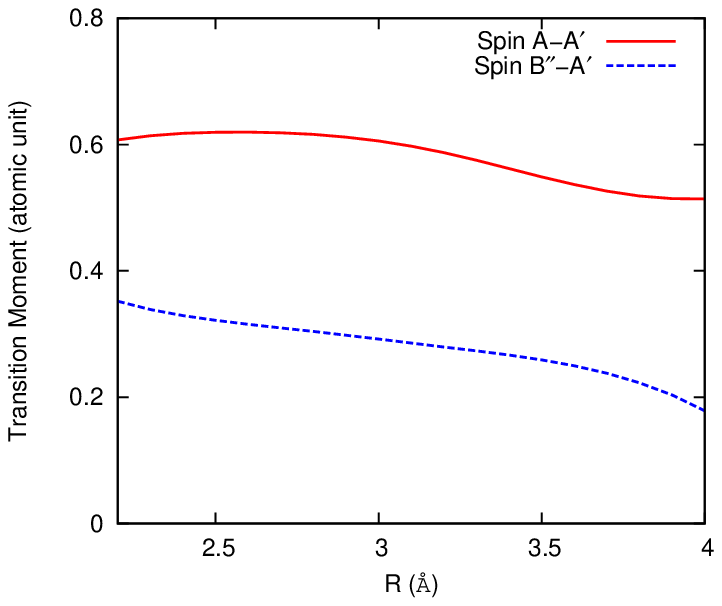}%
 \caption{
 Left panel: Electric dipole transition moment between the $X$ and $A$ states of I$_2$ (red solid line), 
 and between the $X$ and $B''$ states (blue dashed line). 
 Right panel: Spin transition moment between the $A$ and $A'$ states (red solid line), 
 and between the $B''$ and $A'$ states (blue dashed line). 
 }
 \label{transition_mom_Bdd}
\end{figure}

For the wavepacket calculation on the $B''$ state, the same DVR basis parameters were used. 
Since the $B''$ state is repulsive, we need to put the absorbing potential from $R$ = 5.0 to 7.0 \AA~to prevent artificial reflection of 
the wavepacket at the boundary. This absorbing potential can be regarded as decay of the wavepacket 
into $R = \infty$. 
The wave packet on the $B''$ potential energy curve with the specific initial state $S|A'(v)\rangle$, 
where $A'(v)$ represents the vibrational state on the $A'$ state, is propagated by expanding the time evolution operator in terms of 
Chebyshev polynomials \cite{ISI:A1984TS41400029}, with $\Delta t$ = 0.1 fs. 
The total propagation time is 400 fs, which is much shorter than the time needed in the wave packet propagation on the $A$ electronic state. 
This short propagation time can be attributed to the fact that the wave packet is not reflected, 
but just absorbed in 400 fs at the right-hand side of the $B''$ potential energy curve. 
The diploe and spin transition moments, necessary to evaluate Eq.~(\ref{eq_wp}) are shown in Fig. \ref{transition_mom_Bdd}.

\begin{figure}[h]
 \centering\includegraphics[scale=1.0]{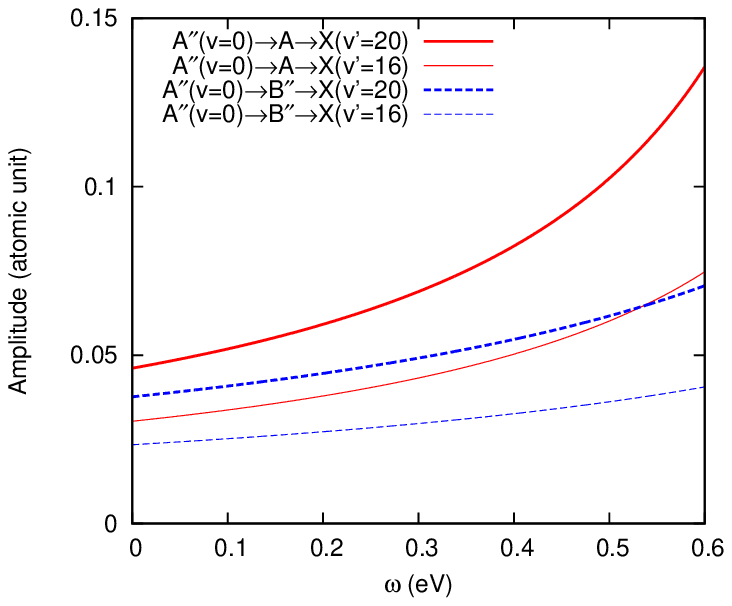}%
  \caption{
Comparison of the amplitudes, Eq.~(\ref{eq_wp}), with the $A1_u$ and $B''1_u$  
intermediate electronic states, obtained by the  wavepacket calculations. 
The initial states are $A'(v=0)$, and the final states are $X(v'=16,20)$. 
}
\label{WP_A_Bdd}
\end{figure}

The amplitudes for the $A' \to X$ process with the $A$ and $B''$ intermediate electronic states, obtained by the 
wavepacket calculations, are compared in Fig.~\ref{WP_A_Bdd}. 
When the initial vibrational state is $A'(v=0)$ and the final vibrational state is $X(v'=20)$, 
the amplitude of the $A$ state at $\omega = 0.4$ eV is about 60 \% larger than that of the $B''$ state. 
This means that the molecular factor including the $A$ intermediate state is approximately 2.5 times larger than 
that including the $B''$ intermediate state. 
Although the contribution of the $B''$ state is not negligible, for the purpose of the present study, 
the molecular factor may be safely approximated by including only the $A$ electronic state.

%

\vfill\pagebreak


\begin{thebibliography}{10}
\bibitem{majorana}
 The massive Majorana neutrino is described by a relativistically 
covariant two-component spinor and
satisfies the Majorana equation (unlike the Dirac equation
for four-component spinors).
The particle number is ill-defined in the Majorana equation, and
violates the U(1) symmetry associated with the lepton number.
The quasi-particle of Majorana nature discussed in
condensed matter physics is different in
its dispersion relation from the massive Majorana neutrino in particle physics.
\bibitem{comparison_with_i2_ptep}
 Due to a non-trivial $\omega$ dependence of the molecular factor $C^a(\omega)$
as calculated in the preceding section,
the overall spectral form shown in Fig.~\ref{i2 overall sepctrum} 
is considerably different from
the one previously given in Ref.~\cite{ptep_overview} where
the precise calculation of  $C^a(\omega)$ is not attempted and the $\omega$
dependence of  $C^a(\omega)$ has been ignored.
\bibitem{fc_factor}
 In Ref.~\cite{ptep_overview} the Franck-Condon factor of I$_2$
was calculated using the Morse potential where relevant
parameters are determined by experimental data.
Our new rate estimate for the transition $A(v=0) \rightarrow 
X(v' = 24)$ gives a value $\sim 0.03$  arising from the Franck-Condon factor.
\end{thebibliography}

\end{document}